\documentstyle[aps,preprint,floats,epsf,epsfig]{revtex}
\begin{document}
\def\be{\begin{eqnarray}}
\def\en{\end{eqnarray}}
\def\non{\nonumber}
\def\la{\langle}
\def\ra{\rangle}
\def\nc{N_c^{\rm eff}}
\def\vp{\varepsilon}
\def\A{{\cal A}}
\def\B{{\cal B}}
\def\up{\uparrow}
\def\dw{\downarrow}
\def\vma{{_{V-A}}}
\def\vpa{{_{V+A}}}
\def\smp{{_{S-P}}}
\def\spp{{_{S+P}}}
\def\J{{J/\psi}}
\def\ov{\overline}
\def\Lqcd{{\Lambda_{\rm QCD}}}
\def\pr{{\sl Phys. Rev.}~}
\def\prl{{\sl Phys. Rev. Lett.}~}
\def\pl{{\sl Phys. Lett.}~}
\def\np{{\sl Nucl. Phys.}~}
\def\zp{{\sl Z. Phys.}~}
\def\lsim{ {\ \lower-1.2pt\vbox{\hbox{\rlap{$<$}\lower5pt\vbox{\hbox{$\sim$}
}}}\ } }
\def\gsim{ {\ \lower-1.2pt\vbox{\hbox{\rlap{$>$}\lower5pt\vbox{\hbox{$\sim$}
}}}\ } }

\font\el=cmbx10 scaled \magstep2{\obeylines\hfill January, 2002}

\vskip 1.5 cm

\centerline{\large\bf Penguin-Induced Radiative Baryonic $B$
Decays}
\bigskip
\centerline{\bf Hai-Yang Cheng$^{1,2}$ and Kwei-Chou Yang$^{3}$}
\medskip
\centerline{$^1$ Institute of Physics, Academia Sinica}
\centerline{Taipei, Taiwan 115, Republic of China}
\medskip
\centerline{$^2$ C.N. Yang Institute for Theoretical Physics,
State University of New York} \centerline{Stony Brook, New York
11794}
\medskip
\centerline{$^3$ Department of Physics, Chung Yuan Christian
University} \centerline{Chung-Li, Taiwan 320, Republic of China}
\bigskip
\bigskip
\centerline{\bf Abstract}
\bigskip
{\small Weak radiative baryonic $B$ decays $B\to\B_1\ov
\B_2\gamma$ mediated by the electromagnetic penguin process $b\to
s\gamma$ have appreciable rates larger than their two-body
counterparts $B\to\B_1\ov \B_2$. The branching ratios for
$B^-\to\Lambda\bar p\gamma$ and $B^-\to\Xi^0\bar\Sigma^-\gamma$
are sizable, falling into the range of $(1\sim 6)\times 10^{-6}$
with the value preferred to be on the large side, and not far from
the bottom baryon radiative decays $\Lambda_b\to\Lambda\gamma$ and
$\Xi_b\to\Xi\gamma$ due to the large short-distance enhancement
for $b\to s\gamma$ penguin transition and the large strong
coupling of the anti-triplet bottom baryons with the $B$ meson and
the light baryon. These penguin-induced radiative baryonic $B$
decay modes should be accessible by $B$ factories.

}

\pagebreak

{\bf 1.}~~Recently we have presented a systematical study of
two-body and three-body charmful and charmless baryonic $B$ decays
\cite{CYBbaryon1,CYBbaryon2}. Branching ratios for charmless
two-body modes are in general very small, typically less than
$10^{-6}$, except for the decays with a $\Delta$ resonance in the
final state.  In contrast, some of charmless three-body final
states in which baryon-antibaryon pair production is accompanied
by a meson have rates larger than their two-body counterparts, for
example, $\Gamma(\ov B^0\to p\bar n\pi^-)>\Gamma(B^-\to n\bar p)$,
$\Gamma(B^-\to p\bar p K^-)>\Gamma(\ov B^0\to p\bar p)$, and
$\Gamma(\ov B^0\to n\bar p\pi^+)>\Gamma(B^-\to n\bar p)$. In
\cite{CYBbaryon2} we have explained why these three-body modes
have branching ratios larger than the corresponding two-body ones.

In this short Letter, we would like to extend our study to the
weak radiative baryonic $B$ decays $B\to\B_1\ov \B_2\gamma$. At a
first sight, it appears that the bremsstrahlung process will lead
to $\Gamma(B\to\B_1\ov \B_2\gamma)\sim {\cal O}(\alpha_{\rm
em})\Gamma(B\to\B_1\ov \B_2)$ with $\alpha_{\rm em}$ being an
electromagnetic fine-structure constant and hence the radiative
baryonic $B$ decay is further suppressed than the two-body
counterpart, making its observation very difficult at the present
level of sensitivity for $B$ factories. However, there is an
important short-distance electromagnetic penguin transition $b\to
s \gamma$. Owing to the large top quark mass, the amplitude of
$b\to s\gamma$ is neither quark mixing nor loop suppressed.
Moreover, it is largely enhanced by QCD corrections. As a
consequence, the short-distance contribution due to the
electromagnetic penguin diagram dominates over the bremsstrahlung.
This phenomenon is quite unique to the bottom hadrons which
contain a heavy $b$ quark; such a magic short-distance enhancement
does not occur in the systems of charmed and strange hadrons. It
has been suggested in \cite{HS} that charmless baryonic $B$ decays
may be more prominent in association with a photon emission. We
shall see that the radiative baryonic $B$ decays proceeded via the
$b\to \gamma$ penguin transition indeed can have appreciable rates
larger than their two-body counterparts.

\vskip 0.3cm {\bf 2.}~~There are several two-body radiative decays
of bottom hadrons proceeding through the electromagnetic penguin
mechanism $b\to s\gamma$:
 \be
 \Lambda_b^0 \to \Sigma^0\gamma,~\Lambda^0\gamma,~~~\Xi^0_b\to\Xi^0\gamma,
~~~\Xi^-_b\to\Xi^-\gamma,~~~\Omega_b^-\to\Omega^-\gamma.
 \en
The radiative baryonic $B$ decays of interest will be
 \be \label{raddecays}
&& B^-\to\{\Lambda\bar p,\,\Sigma^0\bar
p,\,\Sigma^+\bar\Delta^{--},\,\Sigma^-\bar n,
 \,\Xi^0\bar\Sigma^-,\,\Xi^-\bar\Lambda,\,\Xi^-\bar\Sigma^0,\,\Omega^-\bar\Xi^0\}\gamma,
 \non \\ &&
 \ov B^0\to\{\Lambda\bar n,\,\Sigma^0\bar n,\,\Sigma^+\bar p,\,\Sigma^-\bar \Delta^+,
 \,\Xi^0\bar\Lambda,\,\Xi^0\bar\Sigma^0,\,\Xi^-\bar\Sigma^+,\,\Omega^-\bar\Xi^+\}\gamma.
 \en
Note that in our notation, $\bar\Delta^{--}$ means the
antiparticle of $\Delta^{++}$ and $\bar\Sigma^+$ the antiparticle
of $\Sigma^-$.

Let us first consider the decay $B^-\to\Lambda\bar p\gamma$ as an
illustration. The short-distance $b\to s\gamma$ penguin
contribution is depicted in Fig. 1. Since a direct evaluation of
this diagram is difficult, we shall instead evaluate the pole
diagrams shown in Fig. 1. However, the meson $K^*$ pole amplitude
is expected to be suppressed as the intermediate $K^*$ state is
far off mass shell. Consequently, the baryon-baryon-$K^*$ coupling
is subject to a large suppression due to the form-factor effects
at large $q^2$. Therefore, we will focus on the pole diagrams with
the strong process $B^-\to\{\Lambda_b^{(*)},\Sigma_b^{0(*)}\}$
followed by the radiative transition
$\{\Lambda_b^{(*)},\Sigma_b^{0(*)}\}\to\Lambda\gamma$. Let us
consider the $\Lambda_b$ pole first. The corresponding pole
amplitude is then given by
 \be
 A(B^-\to\Lambda\bar p\gamma)=ig_{\Lambda_b\to B^-p}\la\Lambda
 \gamma|{\cal H}_W|\Lambda_b\ra\,{1\over (p_\Lambda+k)^2-m_{\Lambda_b}^2}
 \bar u_{\Lambda_b}\gamma_5 v_{\bar p}.
 \en
The relevant Hamiltonian for the radiative $b\to s\gamma$
transition is
 \be
 {\cal H}_W=-{G_F\over\sqrt{2}}\,V_{ts}^*V_{tb}c_7^{\rm eff}O_7,
 \en
with
 \be
 O_7={e\over 8\pi^2}\,m_b\bar
 s\sigma_{\mu\nu}F^{\mu\nu}(1+\gamma_5)b.
 \en
The effective Wilson coefficient $c_7^{\rm eff}$ includes the
contributions from the QCD penguin operators $O_5$ and $O_6$.
Therefore,
 \be \label{rad}
 \la \Lambda(p_\Lambda)\gamma(\vp,k)|{\cal
 H}_W|\Lambda_b(p_{\Lambda_b})\ra=-i{G_F\over\sqrt{2}}\,{e\over 8\pi^2}
 V_{ts}^*V_{tb}\,2c_7^{\rm eff}
 m_b\vp^\mu k^\nu\la\Lambda|\bar s\sigma_{\mu\nu}(1+\gamma_5)b|\Lambda_b\ra.
 \en

\begin{figure}[tb]
\vspace{-1cm}
\hspace{0cm}\centerline{\psfig{figure=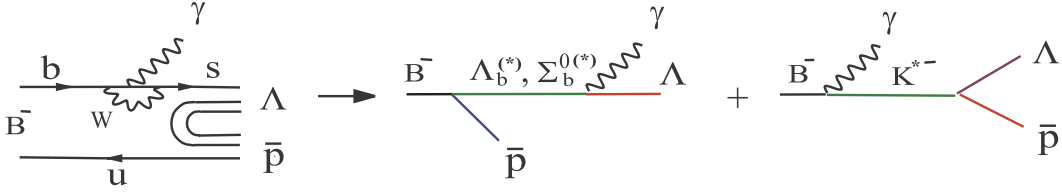,width=15cm}}
\vspace{0cm}
    \caption{{\small Quark and pole diagrams for $B^-\to\Lambda\bar
    p\gamma$.
    }}
\end{figure}

In order to evaluate the tensor matrix elements in Eq.
(\ref{rad}), we consider the static heavy $b$-quark limit so that
$\gamma_0 u_{\Lambda_b}=u_{\Lambda_b}$. This leads to \cite{IW}
 \be
 \la \Lambda|\bar{s}i\sigma_{0i}(1+\gamma_5)b|\Lambda_b\ra=\,\la
\Lambda| \bar{s}\gamma_i(1-\gamma_5)b|\Lambda_b\ra,
 \en
and
 \be
 \la \Lambda|\bar{s}i\sigma_{0i}(1+\gamma_5)b|\Lambda_b\ra \vp^0k^i &=& {1\over 2}\la
\Lambda|\bar{s}\gamma_i(1-\gamma_5)b|\Lambda_b\ra(\vp^0k^i-\vp^ik^0).
 \en
In terms of the baryon form factors defined by
 \be
\la \B_1(p_1)|(V-A)_\mu|\B_2(p_2)\ra &=& \bar
u_1(p_1)\Bigg\{f_1^{\B_1\B_2}(q^2)\gamma_\mu+i{f_2^{\B_1\B_2}(q^2)\over
m_1+m_2} \sigma_{\mu\nu}q^\nu+{f_3^{\B_1\B_2}(q^2)\over
m_1+m_2}q_\mu \non  \\ && -
\Big[g_1^{\B_1\B_2}(q^2)\gamma_\mu+i{g_2^{\B_1\B_2}(q^2)\over
m_1+m_2} \sigma_{\mu\nu}q^\nu+{g_3^{\B_1\B_2}(q^2)\over
m_1+m_2}q_\mu\Big]\gamma_5\Bigg\}u_2(p_2),
 \en
with $q=p_2-p_1$, we obtain
  \be
 \la \Lambda|\bar{s}i\sigma_{0i}(1+\gamma_5)b|\Lambda_b\ra \vp^0k^i
 &=& \bar{u}_\Lambda
 i\sigma_{0i}\vp^0k^i\Big[f^{\Lambda\Lambda_b}_1(0)-f^{\Lambda\Lambda_b}_2(0)
 \non \\
 &+& g^{\Lambda\Lambda_b}_1(0)\gamma_5+{m_{\Lambda_b}-m_\Lambda\over
 m_{\Lambda_b}+m_\Lambda}g_2^{\Lambda\Lambda_b}(0)
 \gamma_5\Big]u_{\Lambda_b}.
 \en
where uses of $\vec{p}_{\Lambda_b}=0$ and the relation
$k^i(\vp^0k^i-\vp^i k^0)=0$ have been made.

Note that there is no contribution from the ${1\over 2}^-$
intermediate state $\Lambda_b^*$ as the matrix element $\la
\Lambda|\bar s(1-\gamma_5)b|\Lambda_b^*\ra$ vanishes. Likewise,
the pole states $\Sigma_b^0$ and $\Sigma_b^{0*}$ also do not
contribute under the factorization approximation because the weak
transition $\la \Lambda|b^\dagger_sb_b|\Sigma_b^{0(*)}\ra$ is
prohibited in the quark model as one can easily check using the
baryon wave functions:
 \be \label{spin-flavor}
\Lambda_b^{\up}&=& {1\over\sqrt{6}}[(bud-bdu)\chi_A+(12)+(13)], \non \\
\Sigma_b^{0 \up}&=& {1\over\sqrt{6}}[(bud+bdu)\chi_s+(12)+(13)],  \\
\Lambda^{\up}&=& {1\over\sqrt{6}}[(sud-sdu)\chi_A+(12)+(13)], \non
 \en
where $abc\chi_s=(2a^\dw b^\up c^\up-a^\up b^\up c^\dw-a^\up b^\dw
c^\up)/ \sqrt{6}$, $abc\chi_A=(a^\up b^\up c^\dw-a^\up b^\dw
c^\up)/\sqrt{2}$,  and $(ij)$ means permutation for the quark in
place $i$ with the quark in place $j$. Hence, only the
intermediate state $\Lambda_b$ makes contributions to the pole
amplitude.

Putting everything together, the pole amplitude reads
 \be \label{poleamp}
 A(B^-\to\Lambda\bar p\gamma)=-ig_{\Lambda_b\to B^-p}\,\bar
 u_\Lambda(a+b\gamma_5)\sigma_{\mu\nu}\vp^\mu k^\nu
 { p\!\!\!/_\Lambda+k\!\!\!/+m_{\Lambda_b}\over
 (p_\Lambda+k)^2-m_{\Lambda_b}^2}\gamma_5 v_{\bar p},
 \en
with
 \be
 a &=& {G_F\over \sqrt{2}}\,{e\over
8\pi^2}2c_7^{\rm eff}m_bV_{tb}V^*_{ts}\,[f_1^{\Lambda\Lambda_b}(0)
-f_2^{\Lambda\Lambda_b}(0)],   \non \\
b &=& {G_F\over \sqrt{2}}\,{e\over 8\pi^2}2c_7^{\rm eff}m_bV_{tb}
V^*_{ts}\,\Big[g_1^{\Lambda\Lambda_b}(0)+{m_{\Lambda_b}-m_\Lambda\over
 m_{\Lambda_b}+m_\Lambda}g_2^{\Lambda\Lambda_b}(0)\Big].
 \en
For the heavy-light form factors $f_i^{\Lambda\Lambda_b}$ and
$g_i^{\Lambda\Lambda_b}$, we will follow \cite{CT96} to apply the
nonrelativistic quark model to evaluate the weak current-induced
baryon form factors at zero recoil in the rest frame of the heavy
parent baryon, where the quark model is most trustworthy. This
quark model approach has the merit that it is applicable to
heavy-to-heavy and heavy-to-light baryonic transitions at maximum
$q^2$.  Following \cite{Cheng97} we have
 \be \label{Lambdabp}
&&
f_1^{\Lambda\Lambda_b}(q^2_m)=g_1^{\Lambda\Lambda_b}(q^2_m)=0.64,
 \quad
 f_2^{\Lambda\Lambda_b}(q^2_m)=g_3^{\Lambda\Lambda_b}(q^2_m)=-0.31,
 \non \\
&&
f_3^{\Lambda\Lambda_b}(q^2_m)=g_2^{\Lambda\Lambda_b}(q^2_m)=-0.10,
 \en
for $\Lambda_b-\Lambda$ transition at zero recoil
$q_m^2=(m_{\Lambda_b}-m_\Lambda)^2$. Since the calculation for the
$q^2$ dependence of form factors is beyond the scope of the
non-relativistic quark model, we will follow the conventional
practice to assume a pole dominance for the form-factor $q^2$
behavior:
 \be
 f(q^2)=f(q^2_m)\left({1-q^2_m/m^2_V\over 1-q^2/m_V^2} \right)^n\,,\qquad
 g(q^2)=g(q^2_m)
\left({1-q^2_m/m^2_A\over 1-q^2/m_A^2} \right)^n\,,
 \en
where $m_V$ ($m_A$) is the pole mass of the vector (axial-vector)
meson with the same quantum number as the current under
consideration. Conventionally, only monopole ($n=1$) and dipole
($n=2$) $q^2$ dependence for baryon form factors is considered. A
recent calculation of the baryon Isgur-Wise function for
$\Lambda_b-\Lambda_c$ transition in \cite{Iva99} indicates that a
monopole $q^2$ dependence is favored. This is further supported by
a recent first observation of $B^-\to p\bar p K^-$ by Belle
\cite{Bellebaryon} which also favors $n=1$ for the momentum
dependence of baryon form factors \cite{CYBbaryon2}.

To compute the branching ratio we shall use the effective Wilson
coefficient $c_7^{\rm eff}(m_b)=-0.31$ (see, e.g. \cite{Beneke}),
the running quark mass $m_b(m_b)=4.4$ GeV and the pole masses
$m_V=5.42$ GeV and $m_A=5.86$ GeV. For quark mixing matrix
elements, we use $|V_{ub}/V_{cb}|=0.085$ and the unitarity angle
$\gamma=60^\circ$. For the strong coupling $g_{\Lambda_b\to B^-
p}$, we note that a fit to the measured branching ratio for the
decay $B^-\to\Lambda_c\bar p\pi^-$ implies a strong coupling
$g_{\Lambda_b\to B^-p}$ with the strength in the vicinity of order
16 \cite{CYBbaryon1}. Hence, we shall use $|g_{\Lambda_b\to B^-
p}|=16$ as a benchmarked value. The pole amplitude (\ref{poleamp})
leads to the branching ratio
 \be
 \B(B^-\to\Lambda\bar p\gamma)=5.9\times 10^{-6}r~~~(0.7\times
 10^{-6}r)
 \en
for $n=1~(n=2)$, where $r\equiv |g_{\Lambda_b\to B^- p}/16|^2$.
The decay rate of $\ov B^0\to\Lambda\bar n\gamma$ is the same as
$B^-\to\Lambda\bar p\gamma$. As noted in passing, the preferred
form-factor momentum dependence is of the monopole form which
implies that $B^-\to\Lambda\bar p\gamma$ has a magnitude as large
as $10^{-5}$. For comparison, we also show the branching ratio for
$\Lambda_b\to\Lambda\gamma$:
 \be
 \B(\Lambda_b\to\Lambda\gamma)=1.9\times 10^{-5}~~~(2.3\times
 10^{-6})
 \en
for $n=1~(n=2)$, which is calculated using the formula
 \be
 \Gamma(\Lambda_b\to \Lambda\gamma)=\,{1\over
8\pi}\left( {m^2_{\Lambda_b}-m^2_\Lambda\over
m_{\Lambda_b}}\right) ^3(|a|^2+|b|^2).
 \en

The weak radiative decay $\Lambda_b\to\Lambda\gamma$ has been
discussed in \cite{Cheng95,CT96,Singer,Mannel,Mohanta} with the
predicted branching ratio spanned in the range of $(0.2-1.5)\times
10^{-5}$. We see that the magnitude of $B^-\to\Lambda\bar p\gamma$
is close to that of $\Lambda_b\to\Lambda\gamma$. In
\cite{CYBbaryon2} we have shown that $2.2\times
10^{-7}<\B(B^-\to\Lambda\bar p)< 4.4\times 10^{-7}$ where the
upper bound corresponds to $\Gamma_{\rm PV}=\Gamma_{\rm PC}$ and
the lower bound to $\Gamma_{\rm PV}=0$, where the subscript PV
(PC) denotes the parity-violating (parity-conserving) decay rate.
Hence, it is safe to conclude that
 \be
 \Gamma(\Lambda_b\to\Lambda\gamma)> \Gamma(B^-\to\Lambda\bar p\gamma)
 >\Gamma(B^-\to\Lambda\bar p).
 \en

Thus far we have not considered next-to-leading order (NLO)
radiative corrections and $1/m_b$ power corrections. It has been
pointed out recently that NLO corrections to $B\to K^*\gamma$
yields an 80\% enhancement of its decay rate \cite{Beneke,Bosch}.
It is thus interesting to see how important the NLO correction is
for the radiative decays $\Lambda_b\to\Lambda\gamma$ and
$B^-\to\Lambda\bar p\gamma$.

\vskip 0.3cm {\bf 3.}~~Let us study other radiative decays listed
in (\ref{raddecays}). For $\ov B\to\Sigma\bar N\gamma$, it
receives contributions from the $\Sigma_b$ pole state, while there
are two intermediate baryon states in the pole diagrams for $\ov
B\to\Xi\ov\B_s\gamma$: the anti-triplet bottom baryon $\Xi_b$ and
the sextet $\Xi_b'$. There are two essential unknown physical
quantities: strong couplings and baryon form factors. For strong
couplings we will follow \cite{Yaouanc,Jarfi} to adopt the $^3P_0$
quark-pair-creation model in which the $q\bar q$ pair is created
from the vacuum with vacuum quantum numbers $^3P_0$. We shall
apply this model to estimate the relative strong coupling strength
and choose $|g_{\Lambda_b\to B^- p}|=16$ as a benchmarked value
for the absolute coupling strength. The results are (for a
calculational detail, see \cite{CYBbaryon2})
 \be \label{gSigb}
 g_{\Lambda_b\to B^- p}=3\sqrt{3}\,g_{\Sigma_b^0\to
 B^-p}=3\sqrt{3}\,g_{\Sigma_b^0\to\ov B^0n}=-{3\sqrt{3\over 2}}\,g_{\Sigma_b^+\to\ov B^0p}=
 -{3\sqrt{3\over 2}}\,g_{\Sigma_b^-\to B^-n},
 \en
and
 \be \label{gXib}
 g_{\Lambda_b\to B^- p} &=& g_{\Xi_b^{0}\to
 B^-\Sigma^+}=g_{\Xi_b^{-}\to \ov
 B^0\Sigma^-}=\sqrt{2}\,g_{\Xi_b^{0}\to\ov B^0\Sigma^0}=\sqrt{2}\,g_{\Xi_b^{-}\to B^-\Sigma^0}
 \non \\
 &=& \sqrt{6}\,g_{\Xi_b^{0}\to\ov B^0\Lambda}=-\sqrt{6}\,g_{\Xi_b^{-}\to
 B^-\Lambda}=- 3\sqrt{2}\,g_{\Xi_b^{'0}\to\ov B^0\Lambda}
=3\sqrt{2}\,g_{\Xi_b^{'-}\to B^-\Lambda}  \\
 &=&  3\sqrt{3}\,g_{\Xi_b^{'0}\to B^-\Sigma^+}
=3\sqrt{3}\,g_{\Xi_b^{'-}\to \ov
B^0\Sigma^-}=3\sqrt{6}\,g_{\Xi_b^{'0}\to \ov B^0\Sigma^0}
=3\sqrt{6}\,g_{\Xi_b^{'-}\to B^-\Sigma^0}. \non
 \en
We thus see that the anti-triplet bottom baryons $\Lambda_b$ and
$\Xi_b$ have larger couplings than the sextet ones $\Sigma_b$ and
$\Xi_b'$. Therefore, the radiative decay $\ov B\to\Sigma\bar
N\gamma$ is suppressed.

Form factors for $\Sigma_b-\Sigma$, $\Xi_b-\Xi$ and $\Xi_b'-\Xi$
transitions at zero recoil can be obtained by using Eq. (22) of
\cite{CT96}. To apply this equation one needs to know the relevant
spin factor $\eta$ and the flavor factor $N_{\B_1\B_2}$ (see
\cite{CT96} for detail). We find $\eta=-{1\over 3}$, 1, $-{1\over
3}$ and $N_{\B_1\B_2}=1/\sqrt{3}$, $1\/\sqrt{2}$ and $1/\sqrt{6}$,
respectively, for above-mentioned three heavy-light baryonic
transitions. The resultant form factors at zero recoil are
exhibited in Table I where we have applied the baryon wave
functions given in Eq. (A1) of \cite{CYBbaryon2}. For bottom
baryons, we use the masses: $m_{\Lambda_b}=5.624$ GeV \cite{PDG},
$m_{\Sigma_b}=5.824$ GeV, $m_{\Xi_b}=5.807$ GeV and
$m_{\Xi_b'}=6.038$ GeV \cite{Jenkins}.

\begin{table}[ht]
\caption{Various baryon form factors at zero recoil.
 }
\begin{center}
\begin{tabular}{l r r r r r r }
Transition & $f_1$ & $f_2$ & $f_3$ & $g_1$ & $g_2$ & $g_3$  \\
\hline $\Sigma_b-\Sigma$ & 1.73 & 2.05 & $-1.70$ & $-0.21$ &
$-0.03$ & $0.11$ \\
$\Xi_b'-\Xi$ & 1.17 & 1.43 & $-1.20$ & $-0.16$ & 0.04 & 0.10 \\
$\Xi_b-\Xi$ & 0.83 & $-0.50$ & $-0.19$ & 0.83 & $-0.19$ & $-0.50$
 \\
\end{tabular}
\end{center}
\end{table}

Repeating the same calculation as before, we obtain
 \be
  \B(B^-\to\Xi^0\bar\Sigma^-\gamma)&=& 5.7\times 10^{-6}r~~~(0.7\times
 10^{-6}r), \non \\
 \B(B^-\to\Xi^-\bar\Lambda\gamma) &=& 1.2\times 10^{-6}r~~~(1.4\times 10^{-7}r), \\
 \B(B^-\to\Sigma^-\bar n\gamma)&=&2.8\times 10^{-8}r~~~(2.4\times
 10^{-9}r), \non
 \en
for monopole (dipole) momentum dependence of baryon form factors.
The decay rates of other modes satisfy the relations
 \be
 && \Gamma(B^-\to\Sigma^-\bar n\gamma)=\Gamma(\ov
 B^0\to\Sigma^+\bar
 p\gamma)= 2\Gamma(B^-\to\Sigma^0\bar p\gamma)=2\Gamma(\ov B^0\to\Sigma^0\bar
 n\gamma), \non \\
 && \Gamma(B^-\to\Xi^0\bar\Sigma^-\gamma)=\Gamma(\ov
 B^0\to\Xi^-\bar\Sigma^+\gamma)=2\Gamma(B^-\to\Xi^-\bar\Sigma^0\gamma)=2\Gamma(\ov
 B^0\to\Xi^0\bar\Sigma^0\gamma),
 \en
and
 \be
 \Gamma(B^-\to\Xi^-\bar\Lambda\gamma)=\Gamma(\ov
 B^0\to\Xi^0\bar\Lambda\gamma).
 \en
As for the radiative decay of the bottom baryon $\Xi_b$, we obtain
 \be
 \Gamma(\Xi_b\to\Xi\gamma)=1.9\times 10^{-17}\,{\rm
 GeV}~~~(2.4\times 10^{-18}\,{\rm GeV})
 \en
for $n=1$ ($n=2$). It follows that
$\Gamma(\Xi_b\to\Xi\gamma)=1.9\,\Gamma(\Lambda_b\to\Lambda\gamma)$.

\vskip 0.3cm {\bf 4.}~~We have shown that weak radiative baryonic
$B$ decays $B\to\B_1\ov \B_2\gamma$ mediated by the
electromagnetic penguin process $b\to s\gamma$ can have rates
larger than their two-body counterparts $B\to\B_1\ov \B_2$. In
particular, the branching ratios for $B^-\to\Lambda\bar p\gamma$
and $B^-\to\Xi^0\bar\Sigma^-\gamma$ are sizable, ranging from
$1\times 10^{-6}$ to $6\times 10^{-6}$ with the value preferred to
be on the large side, and not far from the bottom baryon radiative
decays $\Lambda_b\to\Lambda\gamma$ and $\Xi_b\to\Xi\gamma$. The
penguin-induced radiative baryonic $B$ decays should be detectable
by $B$ factories at the present level of sensitivity. This is
ascribed to the large short-distance enhancement for $b\to
s\gamma$ penguin transition and to the large strong coupling of
the anti-triplet bottom baryons with the $B$ meson and the octet
baryon. We conclude that radiative baryonic $B$ decays are
dominated by the short-distance $b\to s\gamma$ mechanism.

\vskip 2.5cm \acknowledgments One of us (H.Y.C.) wishes to thank
C.N. Yang Institute for Theoretical Physics at SUNY Stony Brook
for its hospitality.  This work was supported in part by the
National Science Council of R.O.C. under Grant Nos.
NSC90-2112-M-001-047 and NSC90-2112-M-033-004.


\end{document}